\documentclass[twocolumn,amsmath,amssymb,prb]{revtex4-1}

\usepackage{graphicx}
\usepackage{dcolumn} 
\usepackage{bm} 
\usepackage{hyperref} 
\usepackage{xcolor}
\usepackage{amsmath}
\usepackage{amssymb}

\allowdisplaybreaks    
\usepackage{multirow}

\begin{document}

\title{Crystallization of fractional charges in a strongly interacting quasi-helical quantum dot}
\author{F. Cavaliere$^{1,2}$, F. M. Gambetta$^{1,2}$, S. Barbarino$^{3}$, and M. Sassetti$^{1,2}$}
\affiliation{\noindent$^{1}$Dipartimento di Fisica, Universit\`{a} di Genova, Via Dodecaneso 33, I-16146 Genova, Italy\\
\noindent$^{2}$ SPIN-CNR, Via Dodecaneso 33, I-16146 Genova, Italy\\
\noindent$^{3}$ NEST, Scuola Normale Superiore \& Istituto Nanoscienze-CNR, I-56126 Pisa, Italy \\
}
\date{\today}


\begin{abstract}
The ground-state electron density of a one-dimensional spin-orbit coupled quantum dot with a Zeeman term and strong electron interaction is studied at the fractional helical liquid points. We show that at fractional filling factors $\nu=(2n+1)^{-1}$ (with $n$ a non-negative integer) the density oscillates with $N_{0}/\nu$ peaks. For $n\geq 1$ a number of peaks larger than the number of electrons $N_{0}$ suggests that a {\em crystal of fractional quasi-particles} with charge $\nu e$ (with $e$ the electron charge) occurs. The reported effect is amenable of verification via transport measurements in charged AFM-coupled dot.\\ 

\end{abstract}

\maketitle

\section{Introduction}
Quasi-helical electrons occur in spin-orbit coupled quantum wires when a magnetic field perpendicular to the spin-orbit field is applied~\cite{qh1,qh2,qh3,qh4,qh5,gambprb}. The magnetic field opens a gap at the degeneracy point of the spectrum, when $k_{\mathrm{F}}=k_{\mathrm{SO}}$, with $k_{\mathrm{F}}$ the Fermi wavevector and $k_{\mathrm{SO}}$ the wavevector associated with the spin-orbit interaction. For not too strong magnetic induction, the system is essentially helical with right (left) moving electrons having spin down (up).\\
These systems, object of recent intense theoretical investigations, possess several unusual properties such as the occurrence of Majorana bound states at their edges when proximized with a superconductor~\cite{maj1,maj2} or peculiar spin oscillations in the presence of magnetic impurities~\cite{magin}. Most striking are the effects induced by repulsive electron interactions: as an example, the gap induced by the magnetic field is strongly enhanced by interactions and anisotropic spin properties occur~\cite{qh1,loss1,lk1}.\\
\noindent Strong interactions generate even more interesting effects in quasi-helical wires, when the Fermi level is tuned {\em below} the degeneracy point. A model originally introduced to study fractional gapped phases in arrays of parallel quantum wires~\cite{mw1,mw2} treated as Luttinger liquids~\cite{ll1,ll2,ll3} has been recently applied to the four-channels Luttinger liquid describing spin-orbit coupled quantum wires~\cite{oreg}. The combined effects of the magnetic field and electron-electron scattering are captured by a non-linear Zeeman coupling dressed by interactions~\cite{loss3}. Critical positions of the dot Fermi level are very relevant for fermions, characterized by a {\em fractional} filling factor with {\em odd} denominator
\begin{equation*}
\nu=\frac{k_{\mathrm{F}}}{k_{\mathrm{SO}}}=\frac{1}{2n+1}\quad(n\in\mathbb{N}^{*})\, .
\end{equation*}
Here, interaction becomes resonant and for sufficiently strong interactions it is relevant in the sense of the renormalization group~\cite{oreg,loss3}. As a consequence, additional gaps open and {\em fractional} phases stabilize in the wire. They are accompanied by several signatures, such as the occurrence of low-energy excitations with fractional charge~\cite{oreg,loss3} which induce a non-integer quantization of the conductance~\cite{oreg,loss3,lk2}. Also the shot noise becomes fractionalized~\cite{sela} and unusual properties of the chiral currents have been reported very recently~\cite{sela2}.\\

\noindent What happens to ground-state properties such as the electron density of a quasi-helical {\em quantum dot} when a fractional phase occurs? This is a non-trivial question, since it can be expected that such a peculiar state may exhibit unusual features, which would compete with the conventional Friedel~\cite{giuliani,fabrizio} and Wigner~\cite{giuliani,jauregui,wig1,wig2,wig3,wig4} oscillations. In any {\em finite-size} one-dimensional system, such as a quantum dot embedded into a quantum wire, electron density oscillations occur. A competition exists between oscillations with $2k_{\mathrm{F}}$ wavevector (Friedel oscillations), due to reflections at the dot edges, and oscillations with wavevector $4k_{\mathrm{F}}$ (Wigner oscillations) due to strong electron repulsion~\cite{footnote0}. The prevalence of either of the two is in general controlled by the ratio between the average electron repulsion and kinetic energy. In a dot with one charge and one spin degree of freedom, Friedel oscillations have $N_{0}/2$ peaks (with $N_0$ the number of electrons, assumed even for simplicity) while Wigner oscillations display $N_{0}$ peaks. In quasi-helical systems, Friedel oscillations in the main gap (an integer phase with $\nu=1$) are predicted to display $N_0$ peaks due to the locking between spin and momentum which halves the number of degrees of freedom~\cite{gambepl}.\\

\noindent In this paper, we address this point evaluating the electron density of a quasi-helical quantum dot with open boundary conditions in a fractional phase $\nu=(2n+1)^{-1}$. Along with the standard Friedel and Wigner terms and their higher harmonics~\cite{llwig3}, we consider contributions originating from the dressed Zeeman interactions. We then evaluate the ground-state average electron density.\\

\noindent Our main result is that in the fractional phase $\nu=(2n+1)^{-1}$ the most relevant oscillations of the electron density have wavevector $4(2n+1)k_{\mathrm{F}}$ and exhibit $(2n+1)N_{0}$ peaks. This is in striking contrast with the standard situation where for strong interactions $N_{0}$ peaks occur, namely one for each electron. The presence of a {\em larger} number of peaks than the number of electrons suggests to interpret this result as the manifestation, in the ground state, of a {\em crystal of fractional quasi-particles} with each density peak corresponding to a {\em fractional} charge $\nu e$, with $e$ the electron charge.
\noindent Charge density oscillations can be experimentally detected, for instance in a linear transport experiment performed while scanning the dot with a charged AFM tip~\cite{wig3,wig4,AFM3}. We estimate the conditions to observe this effect in state-of-the-art samples~\cite{InAs1,InAs2,InAs3}. A measurement of such fractional density oscillations would then constitute a probe of the fractional states.\\

\noindent The paper is organized as follows. In Sec.~\ref{sec:the_model} we introduce the Luttinger model of the dot with spin-orbit coupling and magnetic field. Focusing on odd-denominator resonances, we derive an effective Hamiltonian in the strong interactions regime and discuss the electron density operator. In Sec.~\ref{sec:results} our results are reported and discussed in details. Section~\ref{sec:conclusions} contains our conclusions.
\section{The Model}
\label{sec:the_model}
\subsection{Zero magnetic field}
\label{sec:zmf}
We consider a one-dimensional quantum dot of length $\ell$ with open boundaries at $x=0$ and $x=\ell$. The dot is subject to a spin-orbit field acting along the negative direction of the $z$ axis. The single-particle Hamiltonian is~\cite{qh1,gambprb}
\begin{equation}
H_{\mathrm{sp}}=-\frac{1}{2m^{*}}\partial_{x}^2\bm{I}+i g_{\mathrm{SO}}\partial_{x}\bm{\sigma}_{z}	\, ,
\end{equation}
where $m^{*}$ is the effective electron mass, $g_{\mathrm{SO}}$ the strength of the spin-orbit coupling, with associated wavevector $k_{\mathrm{SO}}=m^{*}g_{\mathrm{SO}}$, and $\bm{I}$, $\bm{\sigma}_{z}$ the identity and $z$ Pauli matrix respectively (here and henceforth $ \hbar=1 $).\\
\noindent Including forward electron-electron interactions, the Hamiltonian around the Fermi energy can be described as a two-channels Luttinger liquid
\begin{equation}
\label{eq:H0}
H_{0}=\frac{1}{4\pi}\int_{0}^{\ell}\mathrm{d}x\ \sum_{\mu=\rho,\sigma}v_{\mu}\left[\frac{1}{g_{\mu}}\left(\partial_x \phi_{\mu}\right)^2+g_{\mu}\left(\partial_{x}\theta_{\mu}\right)^2\right]
\end{equation}
with $\mu=\rho\ (\sigma)$ the charge (spin) modes and $v_{\mu}$ their group velocity. The parameter $g_{\mu}$ controls interactions and $v_{\mu}=v_{\mathrm{F}}/g_{\mu}$ in the $\mu$ sector, with $v_{\mathrm{F}}$ the Fermi velocity. For repulsive interactions one has $0<g_{\rho}\equiv g<1$, while $g_{\sigma}=1$ when SU(2) invariance holds. The fields $\phi_{\mu}(x)$, $\theta_{\mu}(x)$ satisfy canonical commutation rules  $\left[\phi_{\mu}(x),\partial_{x'}\theta_{\mu'}(x')\right]=i\pi\delta_{\mu,\mu'}\delta(x-x')$ and can be represented as
\begin{eqnarray*}
\phi_{\mu}(x)&=&i\sqrt{g_{\nu}}\sum_{n_q>0}\frac{e^{-\pi n_{q}\alpha/2\ell}}{\sqrt{n_q}}\sin\left(\frac{\pi n_{q}x}{\ell}\right)\left(b^{\dagger}_{\mu,n_{q}}-b_{\mu,n_{q}}\right)\, ,\label{eq:phi}\\	
\theta_{\mu}(x)&=&\frac{1}{\sqrt{g_{\mu}}}\sum_{n_q>0}\frac{e^{-\pi n_{q}\alpha/2\ell}}{\sqrt{n_q}}\cos\left(\frac{\pi n_{q}x}{\ell}\right)\left(b^{\dagger}_{\mu,n_{q}}+b_{\mu,n_{q}}\right)\, ,\label{eq:theta}		
\end{eqnarray*}
where $\alpha$ is a short-length cutoff and $b_{\mu,n_q}$ are canonical bosonic operators. The fermionic operator is expressed in terms of left and right components
\begin{equation}
\label{eq:fieldop}
\Psi_{s}(x)=e^{-isk_{\mathrm{SO}}x}\left[e^{ik_{\mathrm{F}}x}R_{s}(x)+e^{-ik_{\mathrm{F}}x}L_{s}(x)\right]\, ,
\end{equation}
with $s=\pm$ the $z$ direction of the electron spin. Here, $k_{\mathrm{F}}=\pi N_{0}/(2\ell)$ the Fermi wavevector, with $N_{0}$ the reference number of electrons in the Fermi sea, assumed even for simplicity. The left- and right-moving fields $L_{s}(x)$, $R_{s}(x)$ satisfy $L_{s}(x)=-R_{s}(-x)$ with
\begin{equation}
\label{eq:rmover}
R_{s}(x)=-\frac{i F_s}{\sqrt{2\pi\alpha}}e^{i \pi N_{s}x/\ell}e^{i\frac{\Phi_{s}(x)}{\sqrt{2}}}\, ,
\end{equation}
$F_s$ a Klein factor, $N_{s}$  the number of {\em excess} electrons in the $s=\pm$ spin sector with respect to the Fermi sea and 
\begin{equation}
\label{eq:Phi}
\Phi_{s}(x)=\theta_{\rho}(x)+s\theta_{\sigma}(x)-\phi_{\rho}(x)-s\phi_{\sigma}(x)\, .
\end{equation}
Note that $\left[\Phi_{s}(x),\Phi_{s'}(-x)\right]=8i\delta_{s,s'}f(x)$ with
\begin{equation}
\label{eq:fofx}
f(x)=\frac{1}{2}\arctan\left[\frac{\sin\left(\frac{2\pi x}{\ell}\right)}{e^{\pi\alpha/\ell}-\cos\left(\frac{2\pi x}{\ell}\right)}\right]\, ,
\end{equation}
due to the finite size of the dot~\cite{llwig4}.

\noindent When no excess charges are considered, $N_{s}=0$, the Fermi energy crosses the energy spectrum~\cite{qh1} at the four wavevectors $\pm(k_{\mathrm{F}}\pm k_{\mathrm{SO}})$. 
\subsection{Magnetic field}
\label{sec:mf}
A magnetic field along the $x$ axis induces a Zeeman coupling $V=\frac{1}{2}\mu_Bg^*\bar{B}\bm{\sigma}_{x}\equiv\frac{B}{2}\bm{\sigma}_{x}$ with $\bar{B}$ the field intensity, $ \mu_B $ the Bohr magneton and $\bm{\sigma}_{x}$ the $x$ Pauli matrix. In terms of the fermionic operators
\begin{equation}
V=\frac{B}{2}\int_{0}^{\ell}\mathrm{d}x\ \left[\Psi^{\dagger}_{+}(x)\Psi_{-}(x)+\Psi^{\dagger}_{-}(x)\Psi_{+}(x)\right]\ .	
\end{equation}
When $\Psi_{s}(x)$ is expressed via $L_{s}(x)$ and $R_{s}(x)$ eight terms arise
\begin{equation}
V=\frac{1}{2\pi}\sum_{j=1}^{8}\int_{0}^{\ell}{\mathrm d}x\ V_{j}(x)\, .
\end{equation}
In the presence of electron interactions, each term $V_{j}(x)$ is also {\em dressed} by backscattering interaction vertices~\cite{oreg,loss3} $U_{n}(x)=\left[e^{2ik_{\mathrm{F}}x}L_{\pm}^{\dagger}(x)R_{\pm}(x)\right]^{n}$ as~\cite{footnote}
\begin{equation}
\label{eq:dressing}
V_{j}(x)\to V_{j}^{(n)}(x)=U_{n}(x)V_{j}(x)U_{n}(x)\ ,
\end{equation}
with corresponding Hamiltonian
\begin{equation}
\label{eq:VVV}
V=\frac{1}{2\pi}\sum_{n\geq 0}\sum_{j=1}^{8}\int_{0}^{\ell}{\mathrm d}x\ V_{j}^{(n)}(x)\, .
\end{equation}
In bosonic form,
\begin{equation}
V^{(n)}_{j}(x)=\Delta_{j}^{(n)}\cos\left[2q_{j}^{(n)}x-\sqrt{2}O_{j}^{(n)}(x)\right]\, ,	
\end{equation}
where $\Delta_j^{(n)}$ are interaction amplitudes $\propto B$.
\begin{table}[!ht]
\centering
\begin{tabular}{|l||c|c|}
\cline{1-3}
$\bm{j}$ & $\bm{O_{j}^{(n)}(x)}$  & $\bm{q_{j}^{(n)}}$ \\ \cline{1-3} 
1 & $(\gamma_{n}-1)\phi_{\rho}(x)+\theta_{\sigma}(x)-\phi_{\sigma}(x)$ & $(\gamma_{n}-1)k_{\mathrm{F}}+k_{\mathrm{SO}}$ \\ \cline{1-3}
2 & $(\gamma_{n}-1)\phi_{\rho}(x)+\theta_{\sigma}(x)+\phi_{\sigma}(x)$ & $(\gamma_{n}-1)k_{\mathrm{F}}+k_{\mathrm{SO}}$ \\ \cline{1-3}
3 & $(\gamma_{n}-2)\phi_{\rho}(x)+\theta_{\sigma}(x)$ & $(\gamma_{n}-2)k_{\mathrm{F}}+k_{\mathrm{SO}}$ \\ \cline{1-3}
4 & $\gamma_{n}\phi_{\rho}(x)+\theta_{\sigma}(x)$ & $\gamma_{n}k_{\mathrm{F}}+k_{\mathrm{SO}}$ \\ \cline{1-3}
5 & $(\gamma_{n}-1)\phi_{\rho}(x)-\theta_{\sigma}(x)+\phi_{\sigma}(x)-F_{0}^{(n)}(x)$ & $(\gamma_{n}-1)k_{\mathrm{F}}-k_{\mathrm{SO}}$ \\ \cline{1-3}
6 & $(\gamma_{n}-1)\phi_{\rho}(x)-\theta_{\sigma}(x)-\phi_{\sigma}(x)-F_{0}^{(n)}(x)$ & $(\gamma_{n}-1)k_{\mathrm{F}}-k_{\mathrm{SO}}$ \\ \cline{1-3}
7 & $\gamma_{n}\phi_{\rho}(x)-\theta_{\sigma}(x)-F_{0}^{(n)}(x)$ & $\gamma_{n}k_{\mathrm{F}}-k_{\mathrm{SO}}$ \\ \cline{1-3}
8 & $(\gamma_{n}-2)\phi_{\rho}(x)-\theta_{\sigma}(x)-F_{0}^{(n)}(x)$ & $(\gamma_{n}-2)k_{\mathrm{F}}-k_{\mathrm{SO}}$ \\ \cline{1-3}
\end{tabular}
\caption{Bosonized expression of the eight Zeeman terms, dressed by electron-electron interactions and the corresponding wavevector. Here, $\gamma_{n}=2n+1$ and $F_{0}^{(n)}(x)=2\sqrt{2}(1-\gamma_{n})f(x)$.}
\label{tab:tab1}
\end{table}
The operators $O_{j}^{(n)}(x)$ and wavevectors $q_{j}^{(n)}$ are presented in Tab.~\ref{tab:tab1}, where $\gamma_{n}=2n+1$. We have omitted zero modes describing excess electrons beyond the Fermi sea, which are irrelevant for the forthcoming discussions focusing on the behaviour at the Fermi surface with $N_0$ electrons.\\

\noindent The first four terms $1\leq j\leq 4$ oscillate with a wavelength shorter than $\ell$ and therefore give negligible contributions to the Hamiltonian in the sense of the renormalization group~\cite{oreg,loss3}.\\ 
\noindent On the other hand by suitably tuning $k_{\mathrm{SO}}$ and $k_{\mathrm{F}}$ terms with $5\leq j\leq 8$ can become spatially non-oscillating and thus resonant. In particular, if $k_{\mathrm{F}}=k_{\mathrm{SO}}/(2n)$ ($n\geq 1$), the terms $V_{5}^{(n)}(x)$ and $V_{6}^{(n)}(x)$ become resonant, while for $k_{\mathrm{F}}=k_{\mathrm{SO}}/(2n+1)$ ($n\geq 0$), the terms $V_{7}^{(n)}(x)$ and $V_{8}^{(n+1)}(x)$ resonate. Terms with $j=5,6$ ($j=7,8$) are predicted to give rise to {\em fractional} states in the dot~\cite{mw1,mw2,oreg,loss3} with filling factor $\nu=1/(2n)$ ($\nu=1/(2n+1)$).\\
\noindent The above resonance conditions can be rewritten in terms of the average density $n_{0}=N_{0}/\ell$ as
\begin{equation}
\label{eq:res1}
n_{0}=\frac{2m^{*}g_{\mathrm{SO}}}{\pi\hbar^2}\nu
\end{equation}
where we have re-inserted $\hbar$ for clarity. For a resonance with a given $\nu$, the right-hand side of Eq.~(\ref{eq:res1}) is fixed by the material parameters. By tuning the average particle density via $N_{0}$ and $\ell$ the dot can be brought into resonance. Note that, due to the finite dot length, $k_{\mathrm{F}}$ is quantized so that for a given $n$ and number of electrons $N_0$ only certain values of $k_{\mathrm{SO}}$ satisfy the resonance condition - see also Sec.~\ref{sec:discussion}.\\

\noindent In the following, we will only consider {\em odd denominator} resonances, i.e. those at $ k_{\mathrm{F}}=k_{\mathrm{SO}}/(2n+1) $. Once resonating, the terms $V_{7}^{(n)}(x)$ and $V_{8}^{(n+1)}(x)$ become relevant in the spirit of the renormalization group~\cite{oreg,loss3} when $g\ll g_{\mathrm{c}}$ with $g_{\mathrm{c}}=3/(2n+1)^2$. Thus, for sufficiently strong interactions one can retain only the dominant terms of Eq.~(\ref{eq:VVV})
\begin{eqnarray}
V&\approx &\frac{1}{2\pi}\int_{0}^{\ell}{\mathrm d}x\ \left[V_{7}^{(n)}(x)+V_{8}^{(n+1)}(x)\right]\nonumber\\
&=&\frac{\Delta}{2\pi}\int_{0}^{\ell}{\mathrm d}x\ \cos\left[4f(x)\right]\cos\left[2\eta_{+}(x)\right]\ ,\label{eq:pertodd}
\end{eqnarray}
where $\Delta_{7}^{(n)}=\Delta_{8}^{(n+1)}=\Delta$ and 
\begin{equation}
\label{eq:etap}
\eta_{+}(x)=\frac{1}{\sqrt{2}}\left[\gamma_{n}\phi_{\rho}(x)-\theta_{\sigma}(x)\right]+2\gamma_{n}f(x)\, .	
\end{equation}
\noindent The Hamiltonian $H=H_{0}+V$ can be decoupled into a massless and a massive term, $H\approx h_{0}+h_{M}$, with
\begin{eqnarray}
h_{0}&=&\frac{v_{\eta}}{2\pi}\int_{0}^{\ell}{\mathrm{d}}x\ \left[\frac{1}{g_{\eta}}\left(\partial_{x}\eta_{-}\right)^2+g_{\eta}\left(\partial_{x}\chi_{-}\right)^2\right]\, ,\label{eq:h0}\\	
h_{M}&=&\frac{v_{\eta}}{2\pi}\int_{0}^{\ell}{\mathrm{d}}x\ \left[\frac{1}{g_{\eta}}\left(\partial_{x}\eta_{+}\right)^2+g_{\eta}\left(\partial_{x}\chi_{+}\right)^2\right]+\nonumber\\	
&+&\frac{\Delta}{2\pi}\int_{0}^{\ell}{\mathrm{d}}x\ \cos\left[4f(x)\right]\cos\left[2\eta_{+}(x)\right]\, .\label{eq:hM}
\end{eqnarray}
With the massless field $\eta_{-}(x)=\left[\gamma_{n}\phi_{\rho}(x)+\theta_{\sigma}(x)\right]/\sqrt{2}$ and canonically conjugated $\chi_{\pm}(x)=\left[\gamma_{n}^{-1}\theta_{\rho}(x)\mp\phi_{\sigma}(x)\right]/\sqrt{2}$. Further, we have 
\begin{equation}
\label{eq:prms}
v_{\eta}=\frac{1+\gamma_{n}^2}{4}\frac{v_{\mathrm{F}}}{g_{\eta}}\,;\, g_{\eta}=\gamma_{n}\sqrt{\frac{1+\gamma_{n}^2}{1+\gamma_{n}^2g^2}}\ g\, .
\end{equation}
Note that identifying $H\approx h_{0}+h_{M}$ neglects small corrections $\propto\left(\partial_{x}\eta_{+}\right)\left(\partial_{x}\eta_{-}\right)$ and $\propto\left(\partial_{x}\chi_{+}\right)\left(\partial_{x}\chi_{-}\right)$ which renormalize~\cite{qh1} $v_{\eta}$ and $g_{\eta}$. The massless sector is a standard Luttinger liquid with renormalized parameters~\cite{ll1,ll2,ll3} while the massive term is a sine-Gordon model.\\

\noindent The dynamics of the massless sector is standard. Concerning the massive sector with a sine-Gordon term, the equation of motion for $\eta_{+}(x)$ is
\begin{equation}
\label{eq:eom}
\frac{v_{\eta}}{g_{\eta}}\partial^{2}_{x}\eta_{+}(x)+\Delta\cos\left[4f(x)\right]\sin\left[2\eta_{+}(x)\right]=0\, .
\end{equation}
In the scaling limit $g\ll g_{\mathrm{c}}$ which we consider, $\Delta$ diverges under renormalization group transformations~\cite{oreg,loss3} and the {\em semi-classical} solutions for the equation of motion "pin" the field $\eta_{+}(x)$ to the minima of the cosine term, i.e.
\begin{equation}
\label{eq:scsol}
\eta_{+}(x)=\frac{\pi}{2}+k\pi\quad\quad\quad(k\in\mathbb{Z})\ ,
\end{equation}
with the conjugated field $\chi_{+}(x)$ a strongly fluctuating one: $\langle\chi_{+}^2(x)\rangle\to\infty$ where $\langle\ldots\rangle$ represents the quantum average over the ground state. This is the infinite mass limit of the model~\cite{oreg,loss3}.\\
\noindent When the field $\eta_{+}(x)$ is pinned as in Eq.~(\ref{eq:scsol}) a gap in the spectrum opens. For $n=0$ ($\nu=1$), Eq.~(\ref{eq:pertodd}) is relevant for $g\leq 3$ and corresponds to the opening of a gap near the degeneracy point of the spin-polarized wire subbands. This constitutes the {\em main gap} of the theory, which requires no repulsive interaction among electrons to be formed, where a quasi-helical electron liquid develops~\cite{qh1,qh2,qh3,qh4,qh5,gambprb}.\\ 
\noindent More intriguing are the properties of the {\em fractional} gaps occurring for $n\geq 1$ at fractional fillings
\begin{equation*}
\nu=\frac{1}{2n+1}\, .
\end{equation*}
These gaps have no counterpart in a noninteracting theory and will be the ones addressed in the rest of the paper.\\
\subsection{Electron density operator}
\label{sec:dens}
In order to study the properties of the ground state density, we start by observing that the density operator $\rho(x)$ is composed of several terms~\cite{giuliani,fabrizio,llwig3,llwig1,llwig2,llwig4,llwig5}
\begin{equation}
\label{eq:rho1}
\rho(x)=\rho_{\mathrm{LW}}(x)+\lambda_{\mathrm{F}}\rho_{\mathrm{F}}(x)+\lambda_{\mathrm{W}}\rho_{\mathrm{W}}(x)+\lambda_{\mathrm{Z}}\rho_{\mathrm{Z}}(x)\, ,
\end{equation}
with the weights $\lambda_{i}$ ($i\in\{F,W,Z\}$) free parameters to be suitably determined.\\

\noindent Here, 
\begin{equation}
\label{eq:rholw}
\rho_{\mathrm{LW}}(x)=\frac{N_{0}}{\ell}+\frac{1}{\pi}\partial_{x}\phi_{\rho}(x)\,
\end{equation}
is the long-wave part of the density.\\

\noindent The contribution of the Friedel oscillations~\cite{ll0,giuliani,fabrizio} is
\begin{equation}
\label{eq:rhoF1}
\rho_{\mathrm{F}}(x)=\sum_{m=1}^{\infty}\rho_{\mathrm{F}}^{(m)}(x)
\end{equation}
with
\begin{eqnarray}
\rho_{\mathrm{F}}^{(m)}(x)&=&\cos\left\{2m[k_{\mathrm{F}}x-f(x)]-\sqrt{2}m\phi_{\rho}(x)\right\}\cdot\nonumber\\
&\cdot&\cos\left[\sqrt{2}m\phi_{\sigma}(x)\right]\label{eq:rhoF2}
\end{eqnarray}
the Friedel harmonics at wavevector $2m k_{\mathrm{F}}$ ($m\geq 1$).\\ 

\noindent Interactions induce Wigner oscillations~\cite{llwig3,llwig1,llwig2,llwig4,llwig5} 
\begin{equation}
\label{eq:rhoW1}
\rho_{\mathrm{W}}(x)=\sum_{m=1}^{\infty}\rho_{\mathrm{W}}^{(m)}(x)
\end{equation}
with harmonics
\begin{equation}
\label{eq:rhoW2}
\rho_{\mathrm{W}}^{(m)}(x)=\cos\left\{4m[k_{\mathrm{F}}x-f(x)]-2\sqrt{2}m\phi_{\rho}(x)\right\}
\end{equation}
at wavevector $4m k_{\mathrm{F}}x$.\\

\noindent Finally, a contribution induced by all terms in Eq.~(\ref{eq:VVV}) which do not become resonant 
\begin{equation}
\label{eq:rhoZ1}
\rho_{\mathrm{Z}}(x)=\sum_{m=1}^{\infty}\sum_{j=1}^{8}\left[1-(\delta_{m,n}\delta_{j,7}+\delta_{m,n+1}\delta_{j,8})\right]\rho_{Z,j}^{(m)}\, ,
\end{equation} 
with
\begin{equation}
\label{eq:rhoZ2}
\rho_{\mathrm{Z},j}^{(m)}(x)=\cos\left[2q_{j}^{(m)}x-\sqrt{2}O_{j}^{(m)}(x)\right]\ .
\end{equation}
is considered.\\
These latter terms are motivated by observing that, to the lowest order, an interaction term of the form $V_{j}^{(m)}(x)$ induces a perturbation $\delta\rho_{\mathrm{LW},j}^{(m)}(x)$ to the long-wave term~\cite{llwig5}
\begin{equation*}
\delta\rho_{\mathrm{LW},j}^{(m)}(x)\propto\int_{0}^{\ell}{\mathrm d}x\ \sum_{\mathcal{E}}\frac{\langle 0|V_{j}^{(m)}(x)|\mathcal{E}\rangle\langle\mathcal{E}|\partial_{x}\phi_{\rho}(x)|0\rangle}{E_{\mathcal{E}}-E_{0}}
\end{equation*}
where $|\mathcal{E}\rangle$ represents a many-body excitation of the unperturbed dot with ground state $|0\rangle$ and $E_{\mathcal{E}}$ is the corresponding unperturbed energy. Note that terms with $j=7$, $m=n$ and $j=8$, $m=n+1$ must not be considered here, since they are explicitly included in the Hamiltonian. It can be shown that
\begin{equation}
\label{eq:perturb2}
\delta\rho^{(m)}_{\mathrm{LW},j}(x)\propto\frac{\cos\left[q_{j}^{(m)}x\right]}{q_{j}^{(m)}}
\end{equation}
namely $V_{j}^{(m)}(x)$ produces oscillations with wavevector $q_{j}^{(m)}$. The inclusion of the terms $\rho_{Z,j}^{(m)}(x)$ in the density operator captures the effects of residual interactions not included in the effective Hamiltonian of the system~\cite{ll2,ll3}.\\
\noindent The weights $\lambda_{i}$ ($i\in\{F,W,Z\}$) in Eq.~(\ref{eq:rho1}) are free parameters to be determined imposing suitable constraints on $\rho(x)$, namely open boundary conditions and the conservation of the number of electrons. 
\section{Results}
\label{sec:results}
\subsection{Density oscillations and fractional quasi-particles}
\label{sec:res}
As discussed in Appendix~\ref{sec:app1}, in the infinite mass limit when $\eta_{+}(x)$ is pinned - see Eq.~(\ref{eq:scsol}) - the only terms contributing to the average density $\langle\rho(x)\rangle$ have wavevector $l\cdot4\gamma_{n}k_{\mathrm{F}}$ (with $l\in\mathbb{Z}^{*}$) and scale as $\left[K(x)\right]^{l^2g_{\eta}/2}$ with 
\begin{equation}
K(x)=\frac{\sinh\left(\frac{\pi\alpha}{2\ell}\right)}{\sqrt{\sinh^2\left(\frac{\pi\alpha}{2\ell}\right)+\sin^2\left(\frac{\pi x}{\ell}\right)}}\, ,
\end{equation}
see Eq.~(\ref{eq:K}). When $g\to 0$ (and thus $g_{\eta}\to 0$) the most relevant terms can be selected and the density, takes the final form in Eq.~(\ref{eq:densfin})
\begin{equation}
\langle\rho(x)\rangle\!=\!\frac{N_{0}}{\ell}\left\{1-\frac{\left[K(x)\right]^{\frac{g_{\eta}}{2}}}{5}\sum_{i=1}^{5}\cos\left[4\gamma_{n}k_{\mathrm{F}}x-F_{i}(x)\right]\right\}	,\label{eq:densres}
\end{equation}
where $F_{1}(x)=F_{2}(x)=F_{3}(x)=0$,  $F_{4}(x)=4(\gamma_{n}-1)f(x)$, and $F_{5}(x)=4(\gamma_{n}+1)f(x)$. The terms $F_{4,5}(x)\propto f(x)$ are a direct consequence of the dot finite size~\cite{llwig4} and stem from the terms $\rho_{Z,7}^{(m)}(x)$ and $\rho_{Z,8}^{(m)}(x)$ respectively.\\

\noindent In the main gap $n=0$ (with $\gamma_{n}=1$), the only non-vanishing contributions to $\rho(x)$ are the lowest Wigner harmonic $\rho_{W}^{(1)}(x)$ - see Eq.~(\ref{eq:rhoW2}) - and four terms stemming from $\rho_{Z}(x)$. They all have wavevector $4k_{\mathrm{F}}$. The density displays $N_{0}$ oscillations corresponding to the expected Wigner oscillations of $N_0$ electrons in a box~\cite{qh1}.\\
\noindent For $n\geq 1$, however, the density exhibits $(2n+1)N_{0}$ peaks. This is due to the contribution of the relevant non-zero terms in $\rho_{Z}(x)$ and of the $(2n+1)$-th Wigner harmonic.

\begin{figure}[!ht]
\centering
\includegraphics[width=8cm]{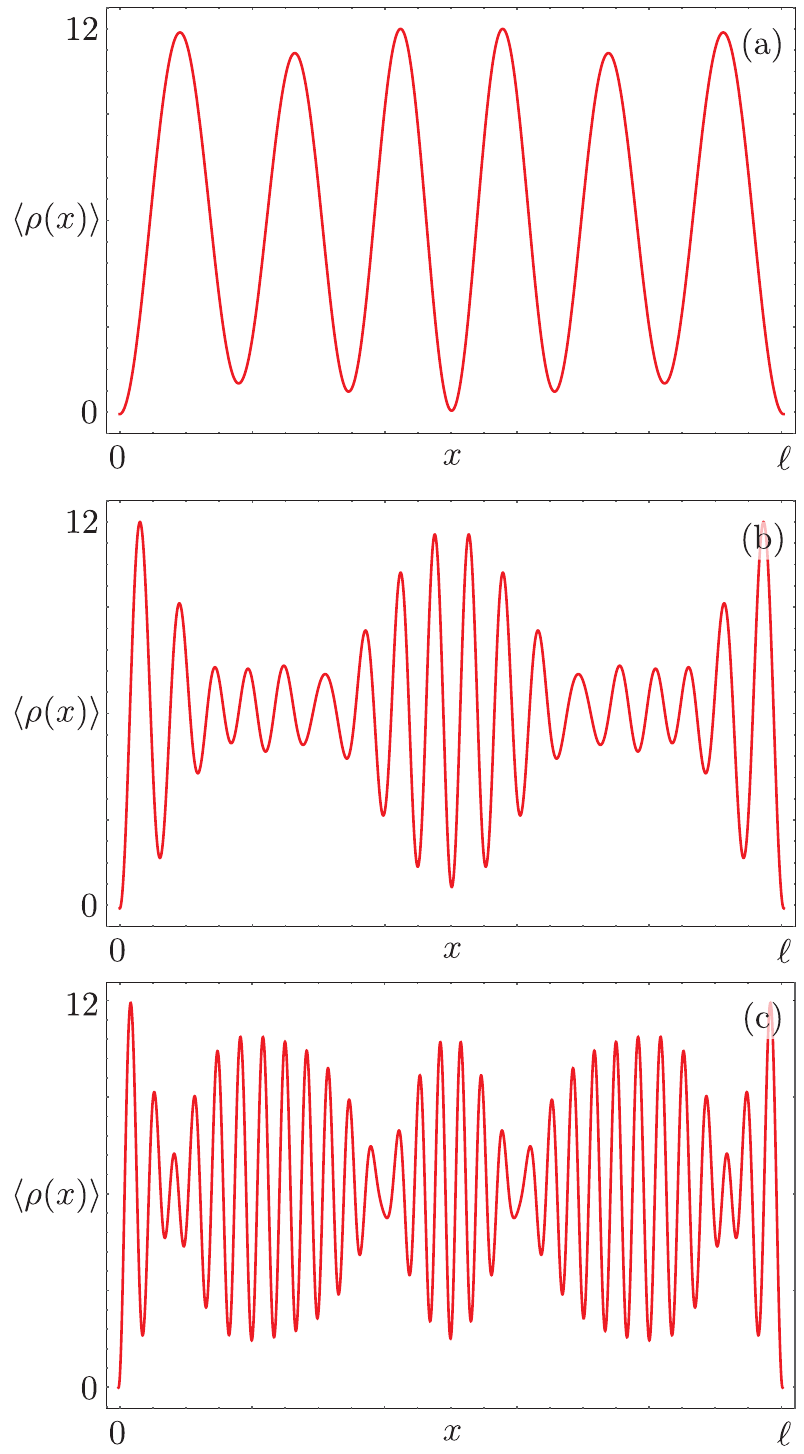}
\caption{\label{fig:fig1}Plot of the $\rho(x)$ (units $\ell^{-1}$) for a dot with $N_0=6$ electrons and $g=0.1$ at the resonance (a) $n=0$; (b) $n=1$; (c) $n=2$. In all panels, $\alpha=k_{\mathrm{F}}^{-1}$.}
\end{figure}

\begin{figure}[!ht]
\centering
\includegraphics[width=8cm]{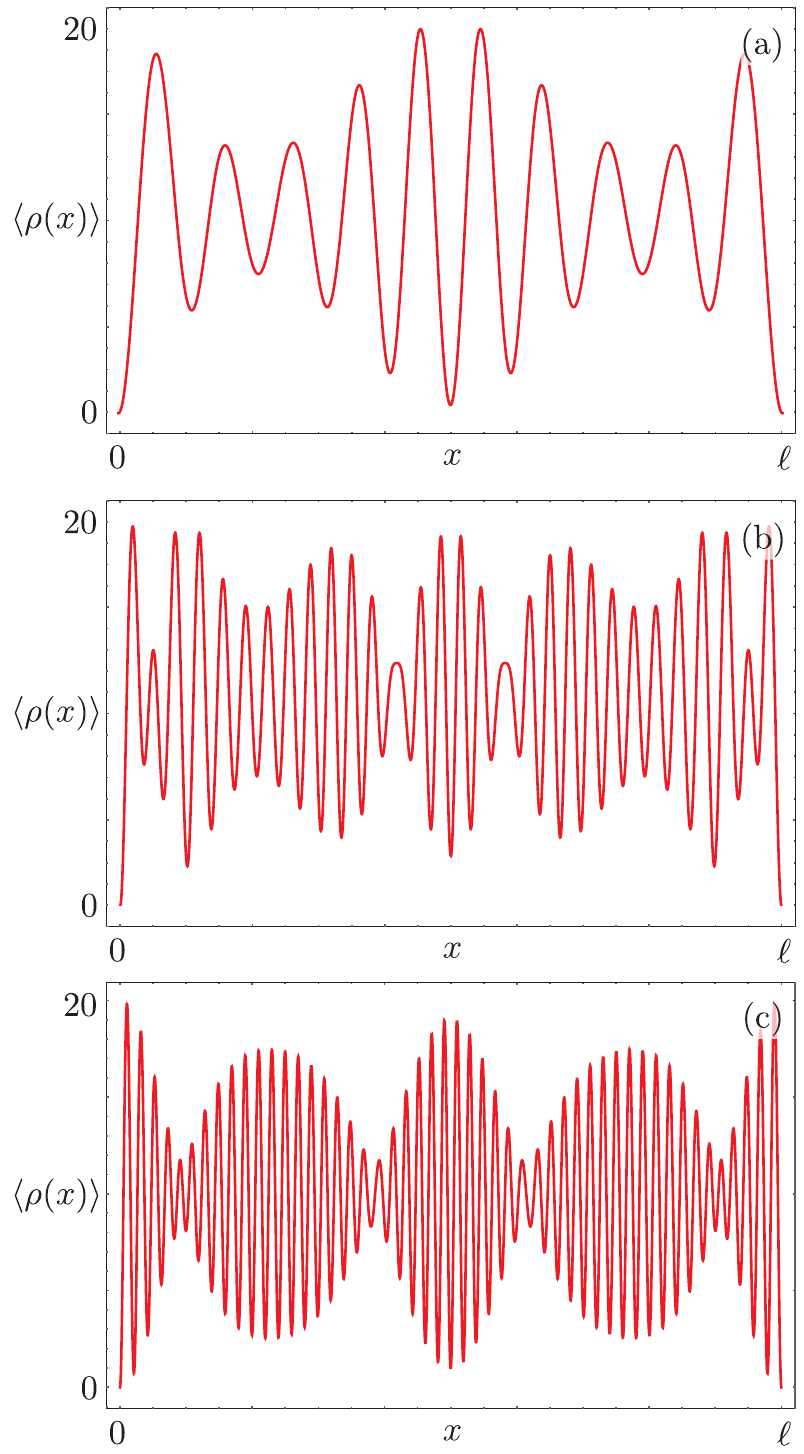}
\caption{\label{fig:fig2}Same as in Fig.~\ref{fig:fig1} but for $N_0=10$. All other parameters as in Fig.~\ref{fig:fig1}.}
\end{figure}

\noindent Figure~\ref{fig:fig1} and Fig.~\ref{fig:fig2} show $\langle\rho(x)\rangle$ for respectively $N_0=6$ and $N_0=10$ particles, in the case of the lowest resonances $n=0$, $n=1$, and $n=2$. As can be seen, the density shows the expected number of peaks, $N_{p}=(2n+1)N_{0}$. Superimposed to the peaks is a slowly-varying modulation induced by the terms $F_{4,5}(x)$ in Eq.~(\ref{eq:densres}). The envelope due to $\left[K(x)\right]^{g_{\eta}/2}$ is also present, but its modulation effect is negligible in the strong interaction regime presented here.\\ 

\noindent For $n\geq 1$, a density displaying $N_{p}>N_{0}$ peaks is a fascinating situation where the many-body wavefunction of the $N_0$ electrons splits up. This splitting can be suggestively interpreted as the crystallization of $(2n+1)N_{0}$ fractional quasi-particles, with fractional charge $e/(2n+1)$, which tend to distribute evenly in space in order to minimize the strong repulsive interactions giving rise to $N_{p}$ distinct peaks in the density.\\
\noindent It is very interesting to observe that, although several predictions have been already made concerning fractional {\em excitations} in strongly interacting spin-orbit coupled wires in the presence of a magnetic field~\cite{oreg,loss3,sela,sela2}, in this work we predict that a bulk ground-state property - the density $\rho(x)$ - exhibits signatures of these fractional quasi-particles.\\

\noindent In this context, very recently a similar phenomenon has been reported in which the density of strongly interacting edge states of a two-dimensional topological insulator with two-particle backscattering exhibits $2N_{0}$ peaks for $N_{0}$ electrons, and has been attributed to the formation of a "fractional Wigner crystal"~\cite{nicoprl}.\\

\noindent We also expect that signatures of fractional quasi-particles show up in higher-order functions, such as in the density-density correlators~\cite{gambepl} (not shown).

\noindent We close this section observing that in deriving Eq.~(\ref{eq:densres}), all the non-zero dressed Zeeman terms $\rho_{Z,j}^{(m)}(x)$ have been weighted with the same factor $\lambda_{Z}$. However, since all the relevant terms have the same wavevector, we expect that the number of oscillations of the density is robust, with only a possibly different modulation pattern of the peaks due to different weights of terms with $F_{4}(x)$ and $F_{5}(x)$. It may be also speculated that within the fractional phase the high-harmonic Wigner oscillations have a weaker amplitude that the term $\rho_{Z}(x)$. Also in this case nonetheless we conclude, in analogy to what discussed previously, that our results are robust barring a slightly different modulation pattern of the peaks.
\subsection{Discussion}
\label{sec:discussion}
We comment here on the observability and stability of the predicted effect.\\

\noindent Assuming InAs as a host material~\cite{InAs1,InAs2,InAs3}, tuning the resonance at $\nu=1/3$ for a dot with $N_{0}=10$ electrons one requires a length~\cite{footnote2} $\ell\approx 7.5\ \mu\mathrm{m}$, while $\ell\approx 4.5\ \mu\mathrm{m}$ is required for $N_{0}=6$ - see Eq.~\ref{eq:res1}. As put forward by several authors, density oscillations in a quantum dot can be probed by scanning a charged AFM tip along the wire while performing a linear transport experiment with lateral source and drain contacts~\cite{wig3,wig4,AFM3,zhuk1,zhuk2}. The tip induces a chemical potential shift proportional to the dot density, and a modulation of the conductance which is connected to the density oscillations~\cite{wig3,wig4,zhuk1,zhuk2}. The lateral resolution can be approximately estimated in the range of~\cite{wig4,zhuk2} $\approx50\ \mathrm{nm}$ and the average wavelength of the density oscillations is $\approx 250\ \mathrm{nm}$. Thus, such an experiment may detect the $N_{p}$ peaks of the density oscillations.\\

\noindent The terms $V_{Z,7}^{(n)}(x)$ and $V_{Z,8}^{(n+1)}(x)$ can become relevant as long as their wavelength $\lambda=2\pi q_{7}^{(n)}$ is larger than the dot length $\ell$, or
\begin{equation}
\label{eq:weakosc}
|\gamma_{n}k_{\mathrm{F}}-k_{\mathrm{SO}}|\leq\frac{2\pi}{\ell}\, .	
\end{equation}
Thus, the number of dot electrons and its length must be calibrated in order to satisfy the above condition. Note that Eq.~(\ref{eq:weakosc}) may allow for some flexibilty for the resonance with $\nu=1/3$: supposing for instance that $\gamma_{n}k_{\mathrm{F}}-k_{\mathrm{SO}}=0$ for $N_0$ dot electrons, one finds that the condition in Eq.~(\ref{eq:weakosc}) is still satisfied for $N_{0}\pm 1$ electrons in the dot. This fact does not hold however for resonances with $n\geq 2$, for which then the tuning of dot parameters is more critical.\\
\noindent The arguments developed in this paper rely on the infinite mass approximation for the sine-Gordon problem, when the gap becomes the largest energy scale. Finite size effects limit the  flow of the renormalization group equations. Therefore, the relevance of the terms $V_{Z,7}^{(n)}(x)$ and $V_{Z,8}^{(n+1)}(x)$ will strongly depend on the bare values of their coupling~\cite{oreg}. The latter, being proportional to the applied magnetic field, should be controllable in experiments~\cite{footnote3}.\\ 

\noindent Renormalization group arguments provide an upper threshold value on the Luttinger parameter $g_{\mathrm{c}}=3/\gamma_{n}^{2}$ in order for a given resonance to be relevant. Estimating the actual strength of electron interactions is however beyond the scope of the model. More insight may be drawn from microscopical approaches such as the Hubbard model. This would not only allow to confirm the predictions made here concerning the fixed point of the theory, but also in principle to observe the transition towards the crystal of fractional quasi-particles. In a recent paper~\cite{sela2} it has been shown that for $V_{Z,7/8}^{(n)}$ to be relevant at $n\geq 1$, on-site repulsion (however strong) is not sufficient in the Hubbard model~\cite{llwig1}. This seems to imply that a non-zero range interaction is needed to observe the predicted effects. Indeed, although the bare Hamiltonian of the dot in Eq.~\ref{eq:H0} includes only zero-range interactions, the correction terms $\rho_{W}(x)$ and $\rho_{Z}(x)$ essentially capture the effects of such long-range interactions.\\
\noindent When interactions are not strong enough, the mass of $\eta_{+}(x)$ does not flow to very large values and additional contributions to the density which are negligible in the regime treated here, may compete or indeed become relevant. In particular, we can expect that lowering interactions the {\em Friedel oscillations} of fractional quasi-particles, with $(2n+1)N_0/2$ peaks, may appear in the density. For even lower interactions electrons would again become relevant and the standard Wigner and Friedel oscillations, respectively with $N_0$ and $N_0/2$ peaks, would set in.\\
\section{Conclusions}
\label{sec:conclusions}
We have studied the density oscillations of a quantum dot created in a spin-orbit coupled quantum wire in the presence of a magnetic field and strong repulsive electron interactions. When the density is tuned at fractional fillings $\nu=k_{\mathrm{F}}/k_{\mathrm{SO}}=1/(2n+1)$, for sufficiently strong interactions a fractional gapped phase develops, for which we have shown that the density oscillates at the peculiar wavevector $4(2n+1)k_{\mathrm{F}}$ and thus shows $(2n+1)N_{0}$ peaks for $N_{0}$ electrons in the dot. We interpret this strong-interaction result as the formation of a crystal of fractional quasi-particles in the ground state of the dot.\\
\noindent We believe that such fractional density oscillations can be detected in linear transport experiments on state-of-the-art samples, probing the dot with a charged AFM tip~\cite{wig3,AFM3}.\\
\begin{acknowledgements}
The authors wish to thank R. Fazio and N. Traverso Ziani for useful discussions.\\

\noindent Financial support of MIUR-FIRB2012, Grant No. RBFR1236VV, is gratefully acknowledged.\\
\end{acknowledgements}
\appendix
\section{Quantum average of the electron density}
\label{sec:app1}
In this Appendix we describe the evaluation of the quantum average of the electron density $\rho(x)$ - see Eq.~(\ref{eq:rho1}) on the ground state in the fractional gapped phase.\\

\noindent The long-wave $\rho_{\mathrm{LW}}(x)$ in Eq.~(\ref{eq:rholw}) term simply evaluates $N_{0}/\ell$.\\

\noindent Each other term - see Eqns.~(\ref{eq:rhoF1},\ref{eq:rhoW1},\ref{eq:rhoZ1}) - can be cast into the form
\begin{equation}
\mathcal{O}=\cos\left[Q(x)+\sum_{p=\pm}[\alpha_{p}\eta_{p}(x)+\beta_{p}\chi_{p}(x)]\right]
\end{equation}
where $Q(x)=qx+c_{Q}f(x)$, with $q$ the wavevector of the oscillating term and $c_{Q}$ a scalar coefficient. To evaluate its average, we convert it in the product form
\begin{equation}
\label{eq:opgen}
\mathcal{O}=e^{iQ'(x)}\prod_{p=\pm}e^{i\alpha_{p}\eta_{p}(x)}\prod_{p=\pm}e^{i\beta_{p}\chi_{p}(x)}+\mathrm{h.c.}\, ,	
\end{equation}
where
\begin{equation}
\label{eq:qprime}
Q'(x)=Q(x)+\alpha_{-}\beta_{+}f(x)	
\end{equation}
also includes the effects of commutations between fields.\\
\noindent Let us first consider the massive sector $p=+$. In the scaling limit of very large mass, $\eta_{+}(x)$ is pinned to the minima of the sine-Gordon potential, see Eq.~(\ref{eq:scsol}). All these minima are degenerate, with the $ j-\text{th} $ minimum having energy $\varepsilon_{j}=-|\varepsilon_{0}|\,,\ \forall j$ with
\begin{equation}
\varepsilon_{0}=-\frac{\Delta}{2\pi}\int_{0}^{\ell}\ \mathrm{d}x\ \cos\left[4f(x)\right]\, .
\end{equation}
The average of $\exp\left[i\alpha_{+}\eta_{+}(x)\right]$ is then
\begin{equation}
\label{eq:avgmass1}
\langle e^{i\alpha_{+}\eta_{+}(x)}\rangle=e^{i\varphi_{0}}\lim_{J\to\infty}\sum_{j=-J}^{J}\frac{e^{-\beta\varepsilon_{j}}e^{i\pi\alpha_{+}j}}{\mathcal{Z}_{J}}	
\end{equation}
where $\varphi_{0}=\pi\alpha_{+}/2$, $\mathcal{Z}_{J}=\sum_{j=-J}^{J}e^{-\beta\varepsilon_{j}}$ with a fictitious small temperature $\propto\beta^{-1}$ which can be set to zero in the end of the calculation. If $\alpha_{+}=2l$, with $l\in\mathbb{Z}$, the above expression reduces to $(-1)^{l}$. If $\alpha_{+}\neq 2l$ however, the partial summations cancel out and the average is zero. Thus we conclude that
\begin{equation}
\label{eq:avgmass2}
\langle e^{i\alpha_{+}\eta_{+}(x)}\rangle=\sum_{l=-\infty}^{\infty}(-1)^{l}\delta_{\alpha_{+},2l}\, .	
\end{equation}
Note that this sharp condition is due to the degeneracy of the minima of the potential (see Sec.~\ref{sec:discussion}).\\
\noindent Furthermore, we have
\begin{equation}
\label{eq:avgmasschi0}
\langle e^{i\beta_{+}\chi_{+}(x)}\rangle=\left[E(x)\right]^{\frac{\beta_{+}^{2}}{2}},
\end{equation}
where
\begin{equation}
\label{eq:avgmasschi}
E(x)=e^{-\langle\chi_{+}^{2}(x)\rangle}\to 0
\end{equation} 
due to the large fluctuations of the field $\chi_{+}(x)$. Thus, every operator which contains the field $\chi_{+}(x) $ will be vanishing in the infinite mass limit.\\

\noindent In the massless sector the (zero temperature) averages $\langle\eta_{-}^{2}(x)\rangle$ and $\langle\chi_{-}^{2}(x)\rangle$ can be evaluated from the correlators
\begin{eqnarray}
\mathcal{G}(x,x';\tau)&=&\langle T_{\tau} \eta_{-}(x,\tau)\eta_{-}(x',0)\rangle\, ,\\
\bar{\mathcal{G}}(x,x';\tau)&=&\langle T_{\tau} \chi_{-}(x,\tau)\chi_{-}(x',0)\rangle\, ,
\end{eqnarray}
where $\tau$ is the imaginary time, as
\begin{eqnarray}
\langle\eta_{-}^{2}(x)\rangle&=&\lim_{\tau\to\alpha/\pi v_{\eta}}\mathcal{G}(x,x;\tau)\\
\langle\chi_{-}^{2}(x)\rangle&=&\lim_{\tau\to\alpha/\pi v_{\eta}}\bar{\mathcal{G}}(x,x;\tau)
\end{eqnarray}
with $\alpha(\pi v_{\eta})^{-1}$ a short imaginary time cutoff. Let us focus on $\mathcal{G}(x,x';\tau)$, as the procedure is analogous for $\bar{\mathcal{G}}(x,x';\tau)$.  Introducing the Fourier transform
\begin{equation}
\mathcal{G}_{\omega}(x,x')=\int_{0}^{\beta}\frac{\mathrm{d}\tau}{2\pi}\ \mathcal{G}(x,x';\tau)e^{-i\omega\tau}\, ,
\end{equation}
from the Hamiltonian in Eq.~(\ref{eq:h0}) one has
\begin{equation}
\left[-\frac{4v_{\eta}}{g_{\eta}}\partial_{x}^{2}+\frac{4\omega^2}{v_{\eta}g_{\eta}}\right]\mathcal{G}_{\omega}(x,x')=\delta(x-x')\, .
\end{equation}
Imposing $\eta_{-}(-x)=-\eta_{-}(x)$ and $\eta_{-}(x+2\ell)=\eta_{-}(x)$ as appropriate for open-boundary conditions, one finds
\begin{equation}
\mathcal{G}_{\omega}(x,x')=\frac{v_{\eta}g_{\eta}}{2\ell}\sum_{j=1}^{\infty}\frac{\sin(q_{j}x)\sin(q_{j}x')}{\omega^2+v_{\eta}^{2}q_{j}^{2}}
\end{equation}
with $q_{j}=(\pi/\ell)j$. Thus, we get
\begin{equation}
\langle\eta_{-}^{2}(x)\rangle=-\frac{g_{\eta}}{4}\log{\left[K(x)\right]}\, ,
\end{equation}
where
\begin{equation}
K(x)=\frac{\sinh\left(\frac{\pi\alpha}{2\ell}\right)}{\sqrt{\sinh^2\left(\frac{\pi\alpha}{2\ell}\right)+\sin^2\left(\frac{\pi x}{\ell}\right)}}\, .\label{eq:K}
\end{equation}
Similarly, one finds
\begin{equation}
\langle\chi_{-}^{2}(x)\rangle=-\frac{1}{4g_{\eta}}\log\left[G(x)\right]\, ,
\end{equation}
with
\begin{equation}
G(x)=8e^{-\frac{\pi\alpha}{\ell}}\sinh^2\left(\frac{\pi\alpha}{2\ell}\right)\cosh\left(\frac{\pi\alpha}{2\ell}\right)K^{-1}(x)\, .\label{eq:G}	
\end{equation}
\noindent Thus in the massless sector one has
\begin{eqnarray}
\exp\left[-\frac{\alpha_{-}^{2}}{2}\langle\eta_{-}^{2}(x)\rangle\right]&=&\left[K(x)\right]^{\alpha_{-}^{2}g_{\eta}/8}\, ,\label{eq:avml1}\\	
\exp\left[-\frac{\beta_{-}^{2}}{2}\langle\chi_{-}^{2}(x)\rangle\right]&=&\left[G(x)\right]^{\beta_{-}^{2}/(8g_{\eta})}\, .\label{eq:avml2}
\end{eqnarray}
\noindent Summarizing, one obtains for the operator in Eq.~(\ref{eq:opgen})
\begin{equation}
\begin{split}
\langle\mathcal{O}\rangle&=\sum_{l}(-1)^{l}\delta_{\alpha_{+},2l}\left[K(x)\right]^{\frac{\alpha_{-}^{2}g_{\eta}}{8}}\left[G(x)\right]^{\frac{\beta_{-}^{2}}{8g_{\eta}}}\\
&\times\left[E(x)\right]^{\frac{\beta_{+}^{2}}{2}}\cos\left[Q'(x)\right]\ .	
\end{split}
\label{eq:avgfinal}
\end{equation}

\noindent The scaling limit imposes severe restrictions to Eq.~(\ref{eq:avgfinal}), due to the simultaneous requirements $\alpha_{+}=2l$, $\beta_{+}=0$, with $l$ an integer.\\ 

\begin{table}[!ht]
\centering
\begin{tabular}{|l||c|c|c|c|c|}
\cline{1-6}
{\bf Term} & ${\bm{\alpha_{+}}}$ & ${\bm{\alpha_{-}}}$ & ${\bm{\beta_{+}}}$ & ${\bm{\beta_{-}}}$ & {\bf Wavevector}\\ \cline{1-6}
$\rho_{F}^{(m)}(x)$ & $-\frac{m}{\gamma_{n}}$ & $-\frac{m}{\gamma_{n}}$ & 1 & -1 & $2mk_{\mathrm{F}}$ \\ \cline{1-6}
$\rho_{W}^{(m)}(x)$ & $-2\frac{m}{\gamma_{n}}$ & $-2\frac{m}{\gamma_{n}}$ & 0 & 0 & $4mk_{\mathrm{F}}$\\ \cline{1-6}
$\rho_{Z,1}^{(m)}(x)$ & $\frac{\gamma_{n}-\gamma_{m}+1}{\gamma_{n}}$ & $\frac{1-\gamma_{m}-\gamma_{n}}{\gamma_{n}}$ & -1 & 1 & $2(\gamma_{m}+\gamma_{n}-1)k_{\mathrm{F}}$ \\ \cline{1-6}
$\rho_{Z,2}^{(m)}(x)$ & $\frac{\gamma_{n}-\gamma_{m}+1}{\gamma_{n}}$ & $\frac{1-\gamma_{m}-\gamma_{n}}{\gamma_{n}}$ & 1 & -1 & $2(\gamma_{m}+\gamma_{n}-1)k_{\mathrm{F}}$\\ \cline{1-6}
$\rho_{Z,3}^{(m)}(x)$ & $\frac{\gamma_{n}-\gamma_{m-1}}{\gamma_{n}}$ & $-\frac{\gamma_{n}+\gamma_{m-1}}{\gamma_{n}}$ & 0 & 0 & $2(\gamma_{m-1}+\gamma_{n})k_{\mathrm{F}}$ \\ \cline{1-6}
$\rho_{Z,4}^{(m)}(x)$ & $\frac{\gamma_{n}-\gamma_{m}}{\gamma_{n}}$ & $-\frac{\gamma_{n}+\gamma_{m}}{\gamma_{n}}$ & 0 & 0 & $2(\gamma_{m}+\gamma_{n})k_{\mathrm{F}}$\\ \cline{1-6}
$\rho_{Z,5}^{(m)}(x)$ & $\frac{1-\gamma_{n}-\gamma_{m}}{\gamma_{n}}$ & $\frac{1-\gamma_{m}+\gamma_{n}}{\gamma_{n}}$ & -1 & 1 & $2(\gamma_{m}-\gamma_{n}-1)k_{\mathrm{F}}$ \\ \cline{1-6}
$\rho_{Z,6}^{(m)}(x)$ & $\frac{1-\gamma_{n}-\gamma_{m}}{\gamma_{n}}$ & $\frac{1-\gamma_{m}+\gamma_{n}}{\gamma_{n}}$ & 1 & -1 & $2(\gamma_{m}-\gamma_{n}-1)k_{\mathrm{F}}$\\ \cline{1-6}
$\rho_{Z,7}^{(m)}(x)$ & $-\frac{\gamma_{n}+\gamma_{m}}{\gamma_{n}}$ & $\frac{\gamma_{n}-\gamma_{m}}{\gamma_{n}}$ & 0 & 0 & $2(\gamma_{m}-\gamma_{n})k_{\mathrm{F}}$\\ \cline{1-6}
$\rho_{Z,8}^{(m)}(x)$ & $-\frac{\gamma_{n}+\gamma_{m-1}}{\gamma_{n}}$ & $\frac{\gamma_{n}-\gamma_{m-1}}{\gamma_{n}}$ & 0 & 0 & $2(\gamma_{m-1}-\gamma_{n})k_{\mathrm{F}}$\\ \cline{1-6}
\end{tabular}
\caption{Coefficients for the different contributions to the electron density.}
\label{tab:tab2}
\end{table}
\noindent Table~\ref{tab:tab2} shows the coefficients and the wavevector for the different oscillating components of the electron density in Eq.~(\ref{eq:rho1}).\\
\noindent It can be seen that $\rho_{F}^{(m)}(x)=\rho_{Z,1}^{(m)}(x)=\rho_{Z,2}^{(m)}(x)=\rho_{Z,5}^{(m)}(x)=\rho_{Z,6}^{(m)}(x)=0$ due to the exponential suppression induced by the large fluctuations of $\chi_{+}(x)$.\\
\noindent The Wigner harmonics only survive if $m=l\gamma_{n}$ (with $l\in\mathbb{Z}^{*}$), see Tab.~\ref{tab:tab2} and Eq.~(\ref{eq:avgmass2}), and have wavevector $l\cdot4\gamma_{n}k_{\mathrm{F}}$. It scales as $\left[K(x)\right]^{l^2 g_{\eta}/2}$. Among them, the most relevant term is $l=1$, with wavevector $Q_0=4\gamma_{n}k_{\mathrm{F}}$.\\
\noindent Going through Tab.~\ref{tab:tab2} one verifies that similar arguments apply for every non-vanishing oscillating term $\rho_{Z,j}^{(m)}(x)$: they all have the wavevector of the form $l\cdot4\gamma_{n}k_{\mathrm{F}}$ (with $l\in\mathbb{Z}^{*}$) and scale as $\left[K(x)\right]^{l^2 g_{\eta}/2}$. Thus, we can conclude that the only contributions to the electron density oscillate with wavevectors $l\cdot4\gamma_{n}k_{\mathrm{F}}$.\\
\noindent Among these terms, the most relevant ones have wavevector $Q_0=4\gamma_{n}k_{\mathrm{F}}$ and scale as $\left[K(x)\right]^{g_{\eta}/2}$. When all the most relevant terms are collected and physical constraints (open boundary conditions and normalization) are imposed on $\langle\rho(x)\rangle$ to determine $\lambda_{W}$ and $\lambda_{Z}$, we obtain
\begin{equation}
\label{eq:densfin}
\langle\rho(x)\rangle\!=\!\frac{N_{0}}{\ell}\left\{1-\frac{\left[K(x)\right]^{\frac{g_{\eta}}{2}}}{5}\sum_{i=1}^{5}\cos\left[Q_{0}x-F_{i}(x)\right]\right\}	,
\end{equation}
where $F_{1}(x)=F_{2}(x)=F_{3}(x)=0$,  $F_{4}(x)=4(\gamma_{n}-1)f(x)$, and $F_{5}(x)=4(\gamma_{n}+1)f(x)$.\\



\begin{thebibliography}{99}
\bibitem{qh1}T. Meng and D. Loss, {\em Phys. Rev. B} {\bf 88}, 035437 (2013).
\bibitem{qh2}P. Streda and P. Seba, {\em Phys. Rev. Lett.} {\bf 90}, 256601 (2003).
\bibitem{qh3}Y. V. Pershin, J. A. Nesteroff, and V. Privman, {\em Phys. Rev. B} {\bf 69}, 121306 (2004).
\bibitem{qh4}C. H. L. Quay, T. L. Hughes, J. A. Sulpizio, L. N. Pfeiffer, K. W. Baldwin, K. W. West, D. Goldhaber-Gordon, and R. de Picciotto, {\em Nat. Phys.} {\bf 6}, 336 (2010).
\bibitem{qh5} T. Meng and D. Loss, {\em Phys. Rev. B} {\bf 87}, 235427 (2013).

\bibitem{gambprb} F. M. Gambetta, N. Traverso Ziani, S. Barbarino, F. Cavaliere, and M. Sassetti, {\em Phys. Rev. B} {\bf 91}, 235421 (2015).

\bibitem{maj1} J. Klinovaja and D. Loss, {\em Phys. Rev. B} {\bf 86}, 085408 (2012).
\bibitem{maj2} D. E. Liu and A. Levchenko, {\em Phys. Rev. B} {\bf 88}, 155315 (2013).
\bibitem{magin} Q. Meng, T. L. Hughes, M. J. Gilbert, and S. Vishveshwara, {\em Phys. Rev. B} {\bf 86}, 155110 (2012).
\bibitem{loss1} B. Braunecker, G. I. Japaridze, J. Klinovaja, and D. Loss, {\em Phys. Rev. B} {\bf 82}, 045127 (2010)
\bibitem{lk1} T. Meng, J. Klinovaja, and D. Loss, {\em Phys. Rev. B} {\bf 89}, 205133 (2014).
\bibitem{mw1} C. L. Kane, R. Mukhopadhyay, and T. C. Lubensky, {\em Phys. Rev. Lett.} {\bf 88}, 036401 (2002).
\bibitem{mw2} J. C. Y. Teo and C. L. Kane, {\em Phys. Rev. B} {\bf 89}, 085101 (2014).
\bibitem{ll0} F. D. M. Haldane, {\em J. Phys. C: Solid State Phys.} {\bf 14}, 2585 (1981).
\bibitem{ll1} J. Voit, {\em Rep. Prog. Phys.} {\bf 58}, 977 (1995).
\bibitem{ll2} J. von Delft and H. Schoeller, {\em Ann. Phys.} {\bf 7}, 225 (1998).
\bibitem{ll3} T. Giamarchi, {\em Quantum Physics in One Dimension} (Oxford Science, Oxford, UK, 2004).
\bibitem{oreg} Y. Oreg, E. Sela, and A. Stern, {\em Phys. Rev. B} {\bf 89}, 115402 (2014).
\bibitem{loss3} T. Meng, L. Fritz, D. Schuricht, and D. Loss, {\em Phys.  Rev. B} {\bf 89}, 045111 (2014).
\bibitem{lk2} D. Rainis, A. Saha, J. Klinovaja, L. Trifunovic, and D. Loss, {\em Phys. Rev. Lett.} {\bf 112}, 196803 (2014).
\bibitem{sela} E. Cornfeld, I. Neder, and E. Sela, {\em Phys. Rev. B} {\bf 91}, 115427 (2015).
\bibitem{sela2} E. Cornfeld, and E. Sela, {\em arXiv:1506.08461v1} (2015).
\bibitem{giuliani} G. Giuliani and G. Vignale, {\em Quantum Theory of the Electron Liquid} (Cambridge University Press, UK, 2005).
\bibitem{fabrizio} M. Fabrizio and A. O. Gogolin, {\em Phys. Rev. B} {\bf 51}, 17827 (1995).
\bibitem{jauregui} K. Jauregui, W. Ha\"usler, and B. Kramer, {\em Europhys. Lett.} {\bf 24}, 581 (1993)
\bibitem{wig1}W. Ha\"usler and B. Kramer, {\em Phys. Rev. B} {\bf 47}, 16353 (1993)
\bibitem{wig2}A. Secchi and M. Rontani, {\em Phys. Rev. B} {\bf 80}, 041404 (2009).
\bibitem{wig3}J. Qian, B. I. Halperin, and E. J. Heller, {\em Phys. Rev. B} {\bf 81}, 125323 (2010).
\bibitem{wig4} N. Traverso Ziani, F. Cavaliere, and M. Sassetti, {\em Phys. Rev. B} {\bf 86}, 125451 (2012). 
\bibitem{footnote0} In principle, both Friedel and Wigner oscillations display a series of higher harmonics at integer multiples of their respective wavevectors, see Ref.~\onlinecite{llwig3}. For standard wires, however, the most relevant contributions is due to the fundamental mode.
\bibitem{gambepl} F. M. Gambetta, N. Traverso Ziani, F. Cavaliere, and M. Sassetti, {\em Europhys. Lett.} {\bf 107}, 47010 (2014).
\bibitem{llwig3} I. Safi and H. J. Schulz, {\em Phys. Rev. B} {\bf 59}, 3040 (1999). 
\bibitem{AFM3} D. Mantelli, F. Cavaliere, and M. Sassetti, {\em J. Phys.: Condens. Matter} {\bf 24}, 432202 (2012).
\bibitem{InAs1} Y. Sidor, B. Partoens, F. M. Peeters, T. Ben, A. Ponce, D. L. Sales, S. I. Molina, D. Fuster, L. Gonz\'alez, and Y. Gonz\'alez, {\em Phys. Rev. B} {\bf 75}, 125120 (2007).
\bibitem{InAs2} S. Csonka, L. Hofstetter, F. Freitag, S. Oberholzer, C. Sch\"onenberger, T. S. Jespersen, M. Aagesen, and J. Nygard, {\em Nano Lett.} {\bf 8}, 3932 (2008).
\bibitem{InAs3} D. Liang, and X. P.A. Gao, {\em Nano Lett.} {\bf 12}, 3263 (2012).
\bibitem{footnote} In principle, more general terms of the form $U_{n}(x)V_{j}(x)U_{n'}(x)$ with $n\neq n'$ could be considered. The results in the regime considered in this paper, however, are not affected by such terms.  
\bibitem{llwig1} J. Schulz, {\em Phys. Rev. Lett.} {\bf 64}, 2831 (1990).
\bibitem{llwig2} J. Schulz, {\em Phys. Rev. Lett.} {\bf 71}, 1864 (1993).
\bibitem{llwig4} Y. Gindikin and V. A. Sablikov, {\em Phys. Rev. B} {\bf 76}, 045122 (2007).
\bibitem{llwig5} A. S\"offing, M. Bortz, I. Schneider, A. Struck, M. Fleischhauer, and S. Eggert, {\em Phys. Rev. B} {\bf 79}, 195114 (2009).
\bibitem{nicoprl} N. Traverso Ziani, F. Cr\'epin, and B. Trauzettel, {\em arXiv:1504.07143} (2015).
\bibitem{footnote2} For this estimate, following Ref.~\onlinecite{InAs2} we have assumed $g_{\mathrm{SO}}=2\cdot10^{-11}\ \mathrm{eV}\cdot\mathrm{m}$, and $m^{*}=0.023\ m_{\mathrm{e}}$ with $m_{\mathrm{e}}=9.1\cdot10^{-31}\ \mathrm{Kg}$ the bare electron mass.
\bibitem{zhuk1} A. A. Zhukov, Ch. Volk, A. Winden, H. Hardtdegen, and Th. Sch\"apers, {\em JETP Lett.} {\bf 142}, 1212 (2012).
\bibitem{zhuk2} A. A. Zhukov, Ch. Volk, A. Winden, H. Hardtdegen, and Th. Sch\"apers, {\em JETP Lett.} {\bf 143}, 158 (2013).
\bibitem{footnote3} Note that, for the case of InAs, a large effective Land\'e $g$-factor is reported~\cite{InAs2}, $g^{*}\approx 5$.
\bibitem{maslov} D. L. Maslov and M. Stone, {\em Phys. Rev. B} {\bf 52}, R5539(R) (1995).
\end{thebibliography}
\end{document}